\documentclass[12pt]{iopart}

\usepackage{iopams}  

\usepackage[numbers,sort&compress]{natbib}
\usepackage{graphicx}
\usepackage{bm}
\usepackage[usenames]{color}
\usepackage{ulem}
\usepackage[caption=false]{subfig}
\usepackage{gensymb}
\usepackage{url}
\usepackage{soul}
\usepackage{multirow}
\usepackage{array}

\graphicspath{{./figures/}}

\newcommand{\strike}[1]{}

\begin{document}

\title[]{Bremsstrahlung emission and plasma characterization driven by moderately relativistic laser-plasma interactions} 

\author{Sushil Singh$^{1,2,3,+}$, Chris D. Armstrong$^4$, Ning Kang$^5$, Lei Ren$^5$, Huiya Liu$^5$, Neng Hua$^5$, Dean R. Rusby$^4$, Ond\v{r}ej Klimo$^{1,6}$, Roberto Versaci$^1$, Yan Zhang$^5$, Mingying Sun$^5$, Baoqiang Zhu$^5$, Anle Lei$^5$, Xiaoping Ouyang$^5$, Livia Lancia$^7$, Alejandro Laso Garcia$^8$, Andreas Wagner$^8$, Thomas Cowan$^8$, Jianqiang Zhu$^5$, Theodor Schlegel$^1$, Stefan Weber$^{1,9}$, Paul McKenna$^{10}$, David Neely$^4$\footnote{deceased}, Vladimir Tikhonchuk$^{1,11}$, Deepak Kumar$^{1,*}$}

\address{$^1$ELI Beamlines, Institute of Physics, Czech Academy of Sciences, 182 21 Prague, Czechia.}
\address{$^2$Department of Radiation and Chemical Physics, Institute of Physics, Czech Academy of Sciences, 182 21 Prague, Czechia.}
\address{$^3$Laser Plasma Department, Institute of Plasma Physics, Czech Academy of Sciences, 182 00 Prague, Czechia.}
\address{$^4$Central Laser Facility, STFC, Rutherford Appleton Laboratory, Didcot OX11 0QX, United Kingdom.}
\address{$^5$National Laboratory on High Power Laser and Physics, Shanghai Institute of Optics and Fine Mechanics, Chinese Academy of Sciences, Shanghai 201800, China.}
\address{$^6$Czech Technical University in Prague, FNSPE, Brehova 7, 115 19 Prague, Czechia.}
\address{$^7$LULI - CNRS, Ecole Polytechnique, CEA: Université Paris-Saclay; UPMC Univ Paris 06: Sorbonne Universities, F-91128, Palaiseau cedex, France.}
\address{$^8$Institute for Radiation Physics, Helmholtz-Zentrum Dresden - Rossendorf, 01328 Dresden, Germany.}
\address{$^9$School of Science, Xi'an Jiaotong University, Xi'an 710049, China}
\address{$^{10}$Department of Physics, SUPA, University of Strathclyde, Glasgow G4 0NG, United Kingdom}
\address{$^{11}$Centre Lasers Intenses et Applications, University of Bordeaux-CNRS-CEA, 33405, Talence cedex, France}

\ead{$^+$singh@ipp.cas.cz; $^*$deepak.kumar@eli-beams.eu}
\vspace{10pt}
\begin{indented}
\item[]September 2020
\end{indented}

\begin{abstract}
Relativistic electrons generated by the interaction of petawatt-class short laser pulses with solid targets can be used to generate bright X-rays via bremsstrahlung. The efficiency of laser energy transfer into these electrons depends on multiple parameters including the focused intensity and pre-plasma level. This paper reports experimental results from the interaction of a high intensity petawatt-class glass laser pulses with solid targets at a maximum intensity of $10^{19}$ W/cm$^2$. {\textit  {In-situ}} measurements of specularly reflected light are used to provide an upper bound of laser absorption and to characterize focused laser intensity, the pre-plasma level and the generation mechanism of second harmonic light. The measured spectrum of electrons and bremsstrahlung radiation provide information about the efficiency of laser energy transfer.
\end{abstract}

\submitto{\NJP}

\section{\label{sec:introduction}Introduction}
High intensity laser pulse interaction with solid targets has many potential applications including fast ignition \cite{Tabak:LLNL:1994}, ion acceleration \cite{Macchi:Pisa:2013} and X-ray generation \cite{Courtois:POP:2011,Courtois:POP:2009,Westover:POP:2014,Cowan:LLNL:2000,Hatchett:LLNL:2000,Norreys:RAL:1999,Schwoerer:Jena:2001}. One of the most fundamental aspects governing these interactions is laser absorption into relativistic electrons. The experiment presented in this paper aims to characterize laser absorption and bremsstrahlung generation on a petawatt-class Nd:glass laser system.

When a laser pulse interacts with a solid target at oblique incidence, a significant fraction of light is reflected in the specular direction from the proximity of the critical density surface i.e., the location where the plasma frequency is equal to the laser frequency. Simultaneously, harmonics of the fundamental laser frequency are generated by either (a) mode conversion from the resonant electric field at the critical density \cite{Erokhin:PTI:1974}, or (b) by coherent wakefield emission \cite{Quere:CEA:2006}, or (c) by reflection from relativistically oscillating critical density surface \cite{Bulanov:GPI:1994,Lichters:MPI:1996}. Due to the different mechanisms of laser absorption and harmonic generation, monitoring the spectrum and intensity of scattered light provides important information about the focused intensity and the laser contrast \cite{Poole:DRACO:2018,Pirozhkov:Astra:2009,Streeter:NJP:2011}. In particular, the curvature of the critical density surface can be inferred from the spatial distribution of the reflected light at the fundamental frequency and comparing the measurements with hydrodynamic simulations of pre-plasma formation and laser absorption.

At the front surface of the target, a significant fraction of the incident laser light is absorbed and strong electromagnetic fields accelerate electrons to relativistic energies. These electrons traverse the target, and subsequently the most energetic electrons escape the rear surface while the remaining are trapped by the sheath potential and re-circulate \cite{Quinn:PPCF:2011}.  The escaped electrons can be directly measured by a magnetic spectrometer while the electrons which are slowed down by collisions within the target are diagnosed by measuring hard X-rays generated by bremsstrahlung \cite{Westover:POP:2014,Singh:RSI:2018}. The multi-MeV hard X-rays generated in such interactions have been investigated by using photo-nuclear activation \cite{Gunther:GSI:2011,Cowan:LLNL:2000,Hatchett:LLNL:2000,Norreys:RAL:1999,Ledingham:Science:2003,Clarke:NIMPRS:2008,Spencer:RSI:2002,Clarke:JRP:2006}. X-ray measurements in the energy range of hundreds of keV to a few MeV are also important as they provide a diagnostic of the fundamental laser-plasma interaction physics, in addition to developing laser based sources for flash radiography applications \cite{Courtois:POP:2009,Courtois:POP:2011,Westover:POP:2014,Galy_2007,Daykin:POP:2018,Sawada:Leopard:2019,Fontaine:ELFIE:2019}. Experiments focused on investigating X-rays in this energy range have revealed the effect of pre-plasma \cite{Courtois:POP:2009} for electron acceleration and concluded that electron re-circulation does not effect the yield of X-rays \cite{Daykin:POP:2018}. The measurements of X-rays in the range of $100$ keV - $1$ MeV, described in this paper confirm that targets with high atomic number (tantalum in this case) and having a thickness of about $2$ mm are optimal for maximizing the X-ray flux of photons with energy $\sim 1$ MeV. Dedicated Monte Carlo simulations performed for the relevant experimental parameters reproduce the dependence of measured X-ray flux on the target thickness.

The paper is organized as follows. The experimental set-up and diagnostics are described in section \ref{sec:experimental-setup}. A description of the experimental results and the related modeling is presented in section \ref{sec:experimental-results}. A brief summary is presented in section \ref{sec:conclusion}.

\section{\label{sec:experimental-setup}Experimental Set-up and Diagnostics}

\begin{figure}
\subfloat[\label{fig:schematic-layout-sgii}]{\includegraphics[width=0.5\textwidth]{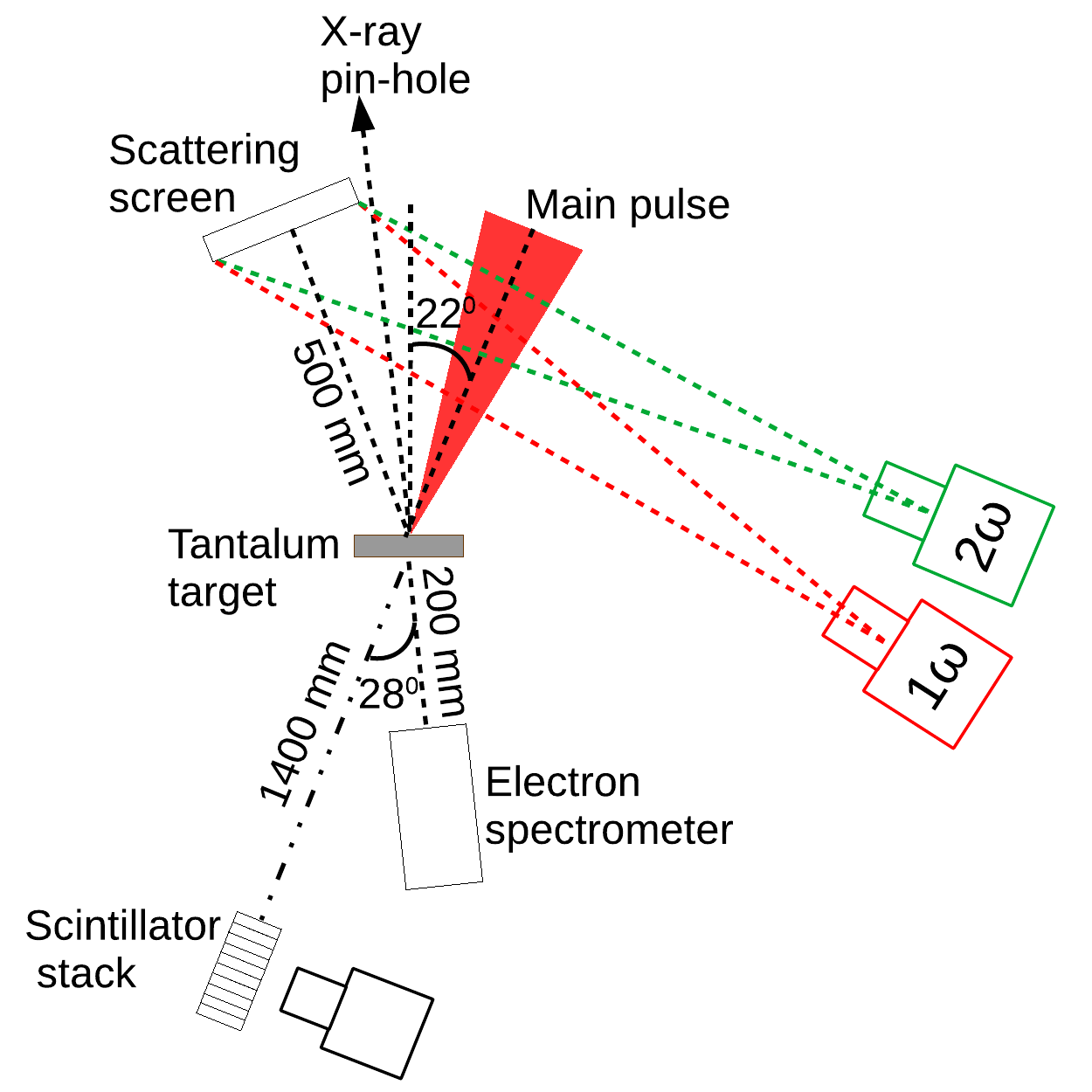}}
\hfill
\subfloat[\label{fig:laser-contrast}]{\includegraphics[width=0.5\textwidth]{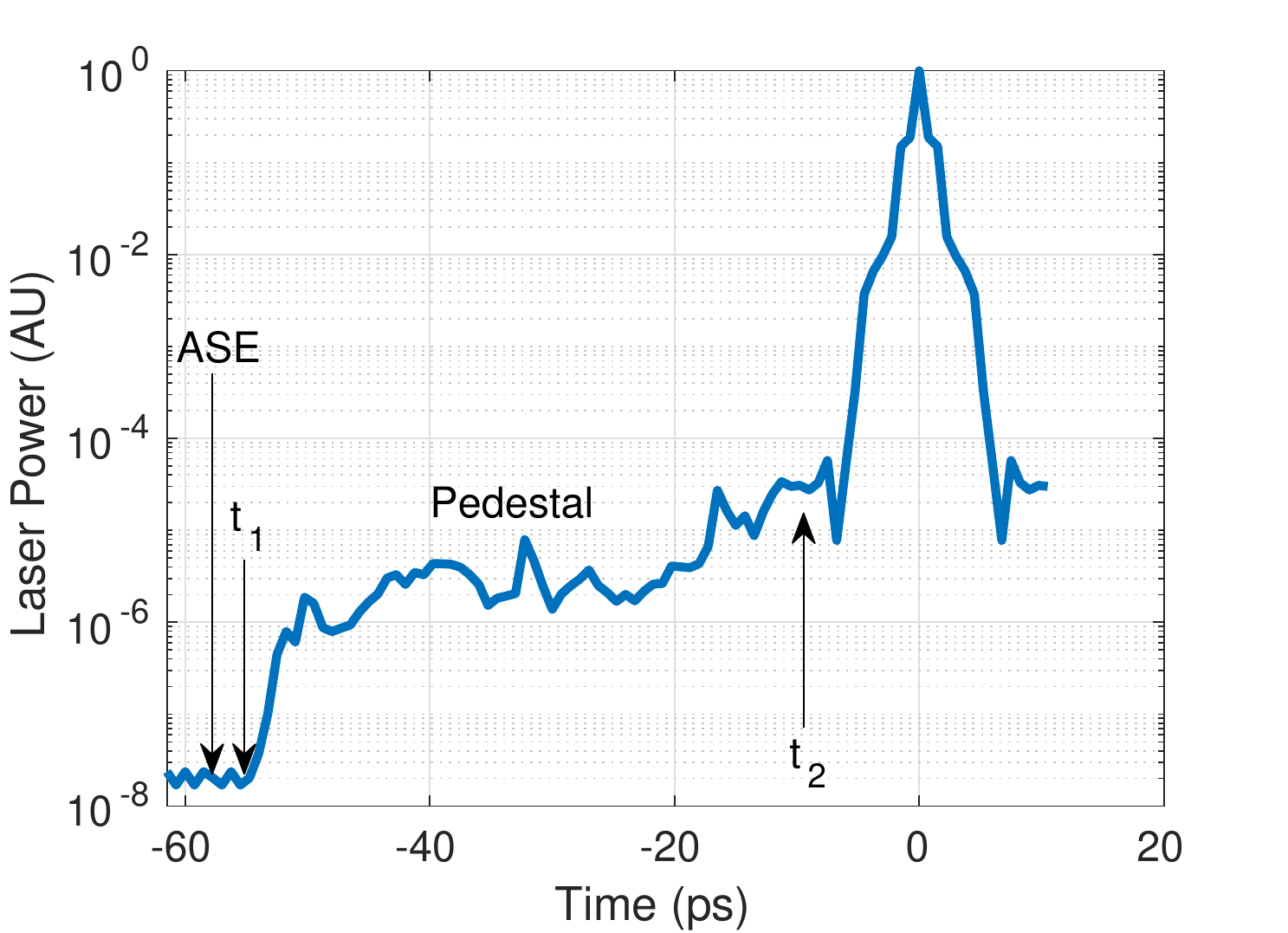}}
\caption{(a) Schematic of the experiment showing the main laser pulse, the imaging system for the scattering screen, the direction of X-ray pin-hole imaging, electron spectrometer and the scintillator stack. The line of sight of the scintillator stack was $18^\circ$ above the horizontal plane (i.e. the plane of the incoming laser beam). (b) Power contrast of the laser beam as measured by a single-shot cross-correlator.}
\end{figure}

The experimental layout at the SG-II upgrade facility \cite{Zhu:SGII:2018} is shown in figure \ref{fig:schematic-layout-sgii}. The main laser pulse was generated by a hybrid optical parametric Chirped Pulse Amplification (CPA) and Nd:glass amplifier. The beam energy was $(300 \pm 25)$ J with a pulse duration of $\approx 1$ ps at a wavelength of $1053$ nm. The p-polarized beam was focused with a f/$2.5$ off-axis parabola (OAP) onto a focal spot of $\approx 45\ \mu$m diameter containing about $80\%$ of the incident energy, thus reaching a peak focused intensity of $10^{19}$ W/cm$^2$ \cite{Zhu:SGII:2018}. The laser intensity contrast was measured using a single-shot cross-correlator with a fiber array and a photo multiplier tube \cite{Ouyang:SGII:2016} and is shown in figure \ref{fig:laser-contrast}. The amplified spontaneous emission (ASE) contrast was $\sim2\times10^{-8}$ and extended till $850$ ps before the main pulse which is half the duration of uncompressed chirped pulse of the laser. The pulse has a contrast pedestal in the range of $10^{-5}$ to $10^{-6}$ for less than $60$ ps before the main pulse.

The focused beam was incident on tantalum (Ta) targets at an angle of $22^\circ$ in the equatorial plane. The thickness of the target was varied in the range from $100\ \mu$m to $4$ mm. To increase the hot electron population, the front surface of some tantalum targets were coated with $10\ \mu$m plastic (parylene) or with glass microspheres of diameter $4-5\ \mu$m. 

A number of diagnostics were used to measure fast electrons, X-rays and the specular reflection of the beam. The accelerated electrons escaping the target were characterised with a magnetic electron spectrometer. It was placed at a distance of $20$ cm from the target at an angle of $28^\circ$ from the laser axis. The $1$ mm entrance aperture of the spectrometer subtended a solid angle of $20\ \mu$sr. A magnetic field of $0.28$ T dispersed the electrons to enable energy resolved detection in the range of $1-35$ MeV. The spectrometer used absolutely calibrated BAS-SR imaging plates as a detector \cite{Singh:RSI:2017,Zeil:RSI:2010} to provide absolute flux of fast electrons from the interaction.

The specular reflection from the target was detected by a scattering screen made of Zenith polymer and having an area of $30$ cm $\times$ $30$ cm. The scattering screen covered a solid angle equivalent to f/$1.67$ on the target, which is greater than the f-number of the focusing parabola f$/2.5$. The scattering screen was imaged using two cameras to monitor the light incident on it at the first and second harmonic of the laser pulse. A band pass interference filter centered at $520$ nm and with a bandwidth of $40$ nm (full width at half maximum) was used with the camera monitoring the second harmonic and a long pass filter with a cut-on wavelength of $1\ \mu$m was used with the camera monitoring the first harmonic. The system was absolutely calibrated using low power continuous wave lasers (at the first and second harmonic) incident on the scattering screen and being imaged by the cameras.

\begin{figure}
  \centering
\begin{tabular}{p{0.5\textwidth} p{0.5\textwidth}}
  \begin{tabular}{p{0.5\textwidth}}
    \subfloat[\label{fig:scintillator-stack-image}]{\includegraphics[width=0.45\textwidth,trim=0cm 0cm 0cm 0cm, clip ]{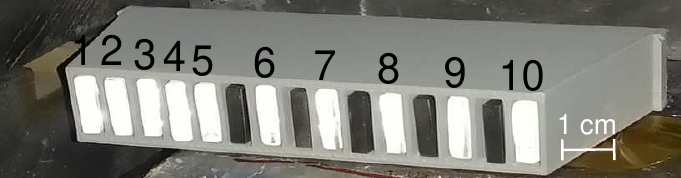}}\\
    \subfloat[\label{fig:scintillator-stack-raw-data-example}]{\includegraphics[width=0.5\textwidth,trim=2.5cm 0cm 2cm 0cm, clip]{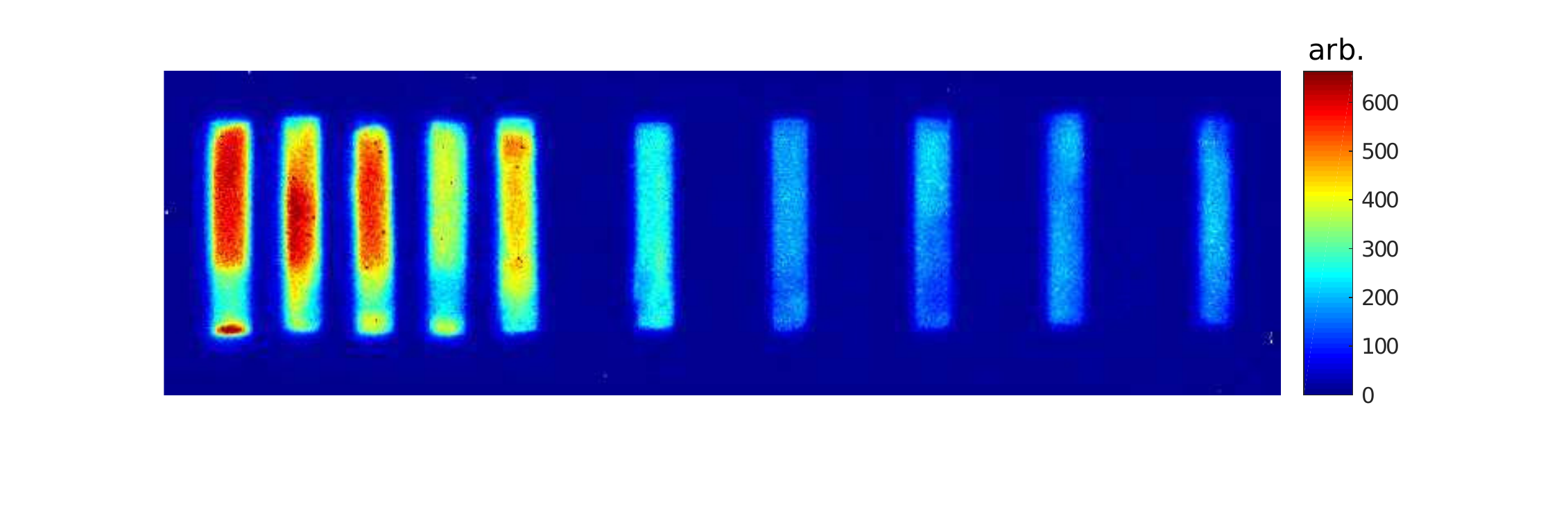}}
  \end{tabular}
  &
  \begin{tabular}{p{0.5\textwidth}}
    \subfloat[\label{fig:scintillator-stack-energy-deposited}]{\includegraphics[width=0.5\textwidth]{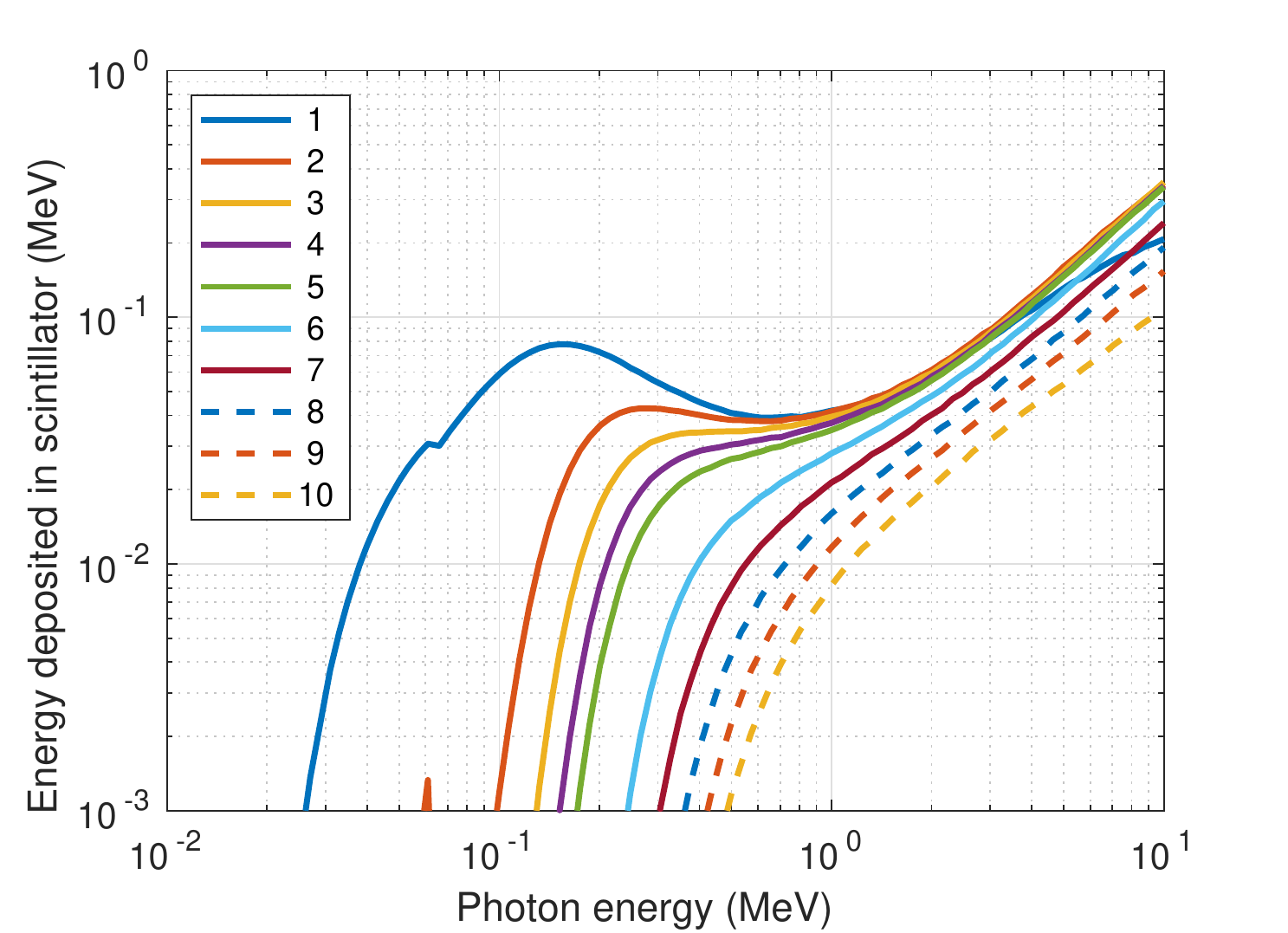}}
  \end{tabular}
\end{tabular}
\caption{\label{fig:scintillator-stack-example} (a) A stack of scintillators and tungsten attenuators placed in a plastic housing and behind a lead collimator. (b) Example of raw data in arbitrary units from imaging the stack of scintillators during a laser shot. (c) Simulated energy deposited in various scintillators in the stack (numbered $1-10$) as a function of incident photon energy.}
\end{figure}

Hard X-rays from $100$ keV to $1$ MeV were measured using a stack of LYSO (Lu$^{1.8}$Y$^{.2}$SiO$^5$:Ce) scintillators \cite{Rusby:RSI:2018,Armstrong:Strathclyde:2019}. Unlike traditional filter stack spectrometers which use passive readouts of image plate \cite{Chen:FilStack:RSI:2008}, this diagnostic provides prompt data from an imaging camera. As shown in figure \ref{fig:schematic-layout-sgii}, the hard X-rays generated by bremsstrahlung were measured at $18^\circ$ above the equatorial plane along the laser axis. The stack of scintillators shown in figure \ref{fig:scintillator-stack-image} was placed outside the vacuum chamber at a distance of $1.4$ m from the target. The X-rays passed through a glass viewport of thickness $1$ cm, and a subsequent permanent magnet of field strength $0.2$ T and with a $2$ cm gap between the poles. The magnet dispersed secondary electrons, preventing them from being incident on the scintillators. Downstream of the magnet, a lead collimator of length $10$ cm, with an aperture of $8\times12$ mm$^2$ was placed to collimate the X-ray beam prior to the scintillators. The aperture subtended a solid angle of $49\ \mu$sr to the source. An example of the data collected from the experiment is shown in figure \ref{fig:scintillator-stack-raw-data-example}. The scintillator stack included ten LYSO crystals and five tungsten filters, each of thickness $2$ mm and cross section $1.1$ cm $\times$ $3$ cm, as can be seen in figure \ref{fig:scintillator-stack-image}. The response of the stack to mono energetic X-rays was simulated using the Monte Carlo code GEANT4 \cite{Chauvie:Geant4:2004,Chauvie:Geant4:2006,Cirrone:Geant4:2010}, and the corresponding transfer matrix for energy deposited in each scintillator is shown in figure \ref{fig:scintillator-stack-energy-deposited}. The stack of scintillators and the camera were covered by black aluminum foil to prevent signal contamination from stray light or laser light. Calibration for the relative efficiency of the individual scintillators in the stack was performed by exposing the stack to a $^{22}$Na radioactive source. The response of the individual scintillators varied within $20\%$.

\begin{figure}
\centering
\includegraphics[width=0.5\textwidth, trim=0.5cm 1.5cm 2.2cm 1cm, clip]{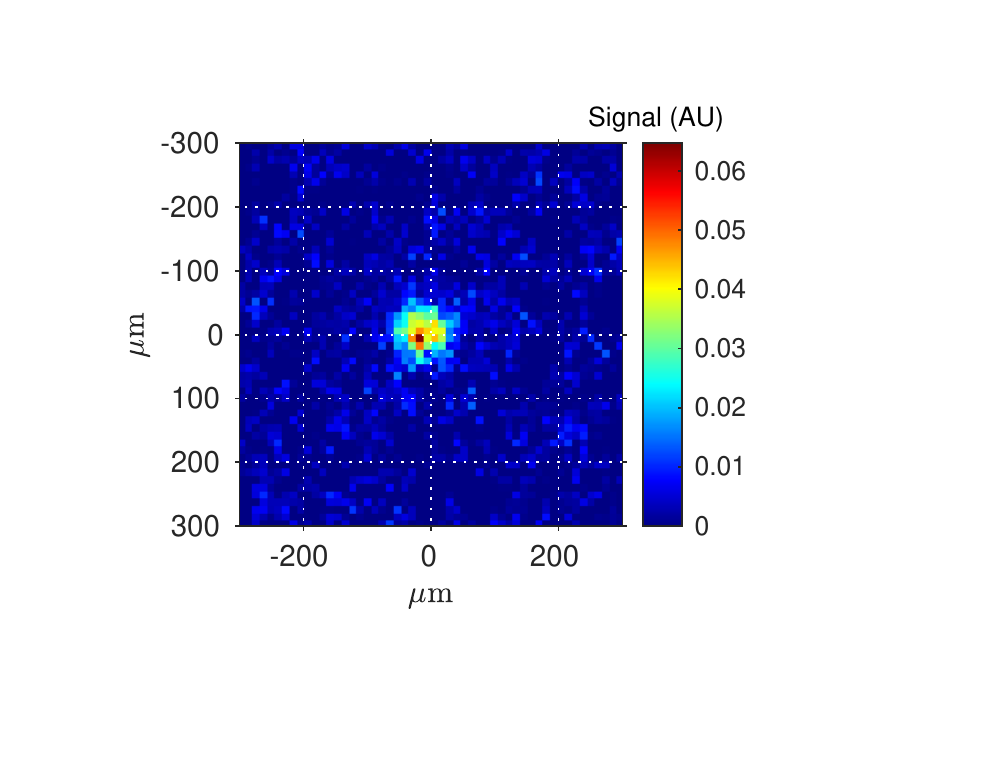}
\caption{\label{fig:focal-spot-measurement} X-ray emission in the range of $0.5-2.5$ keV measured by a pin hole camera from the front surface of the target while shooting a $2$ mm thick Ta target.}
\end{figure}

The interaction of the beam with the target was imaged using a grazing incidence X-ray pinhole camera \cite{Wang:SGIIup:2017}. The camera imaged the front surface of the target and was installed in the direction shown in figure \ref{fig:schematic-layout-sgii}. It imaged soft X-rays in the range of $0.5-2.5$ keV from the target with a magnification of $4.3$. The data is shown in figure \ref{fig:focal-spot-measurement}, and the full width at half max of the signal is $\sim55\ \mu$m which is comparable to the focal spot of the laser.

\section{\label{sec:experimental-results}Experimental Results}

\subsection{\label{sec:specular-reflection}Specular reflection and harmonic generation}

\begin{figure*}
\centering
\includegraphics[width=0.75\textwidth]{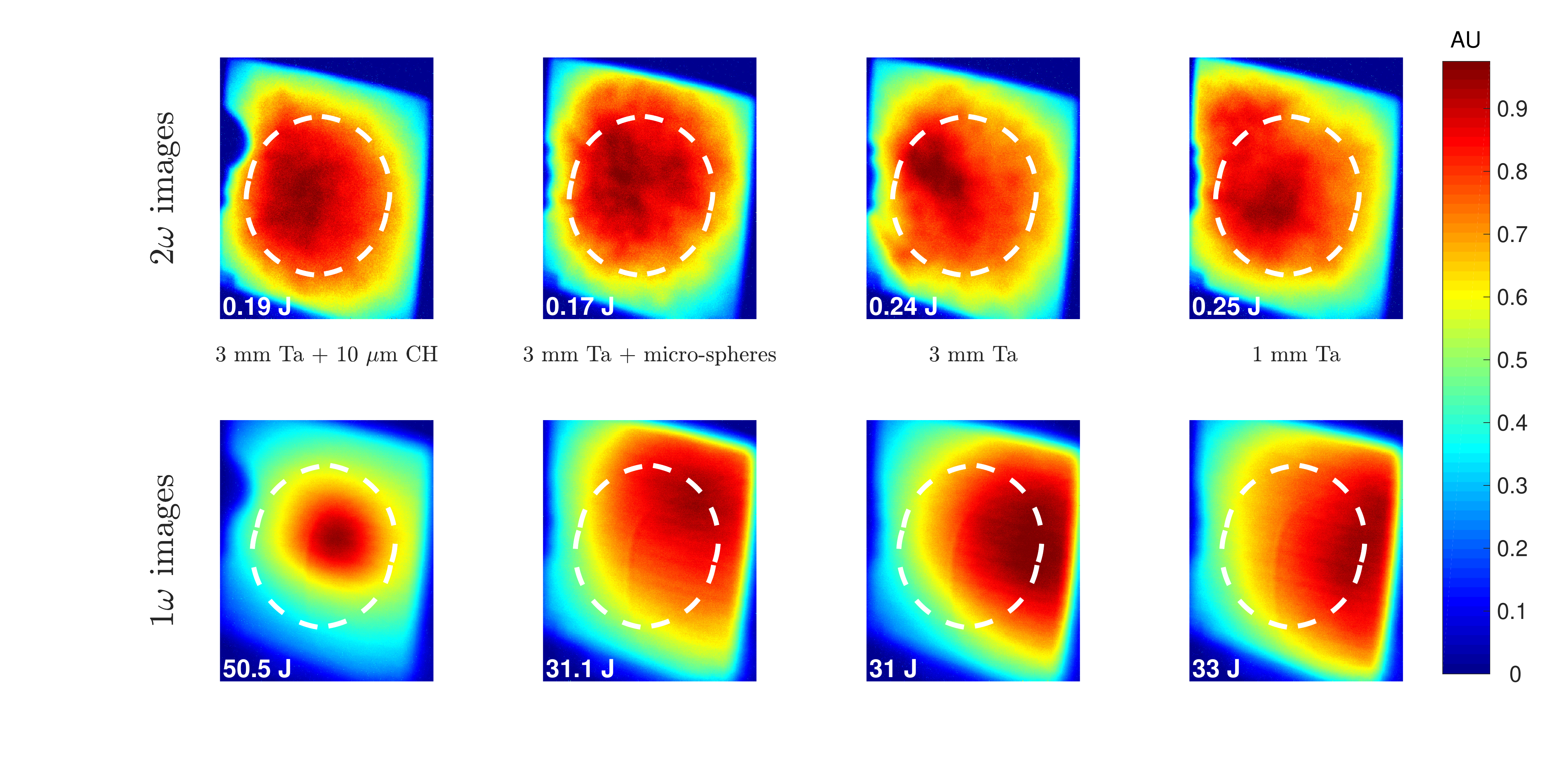}
\caption{\label{fig:results-scattering-1w-2w} Intensity distribution of  the reflected light on the scattering screen at second (top row) and first harmonic (bottom row) of the laser frequency. Each column represents the data collected from a different shot. The target used in each shot is indicated between the rows - $3$ mm thick Ta with $10\ \mu$m thick parylene coating, $3$ mm thick Ta with $4-5\ \mu$m diameter glass microspheres on the front surface, $3$ mm thick Ta, and $1$ mm thick Ta (from left to right). The total energy incident on the scattering screen is indicated at the bottom of each image. The dashed white circle in each image represents the angular aperture of the incident off axis parabola if the target at focus would be an ideal mirror. The white circle appears askew in the images because the shape is compensated for the viewing angle of the cameras.}
\end{figure*}

In order to correlate bremsstrahlung generation with laser coupling at the target front surface, we measured reflected energy in the specular direction. As seen in figure \ref{fig:results-scattering-1w-2w}, the reflection at the fundamental harmonic was affected by the choice of target. For uncoated Ta targets, or for Ta targets covered with $4-5\ \mu$m glass microspheres, the reflection at the fundamental harmonic was diffuse and always shifted to the right, while for plastic coated Ta targets, the reflection was centered along the specular reflection direction. The amount of energy reflected at the fundamental harmonic for plastic coated targets was of the order of $50$ J, which corresponds to about $\sim17\%$ of laser energy. For other targets, the amount of energy at the fundamental harmonic incident on the scattering screen was of the order of $30$ J. However as can be seen from figure \ref{fig:results-scattering-1w-2w}, most of the reflected radiation missed the scattering screen. We estimate that the total amount of reflected laser light could exceed $70$ J, which corresponds to a fraction of $\sim 20-25\%$ of the incident laser energy. This is consistent with measurements on similar laser system by Gray \etal \cite{Gray:GSI:2018}, where about $50\%$ of laser energy was scattered in the first harmonic.

\begin{figure}
\subfloat[\label{fig:scattering-geometry}]{\includegraphics[width=0.5\textwidth, trim=6.5cm 7.5cm 8cm 0cm, clip]{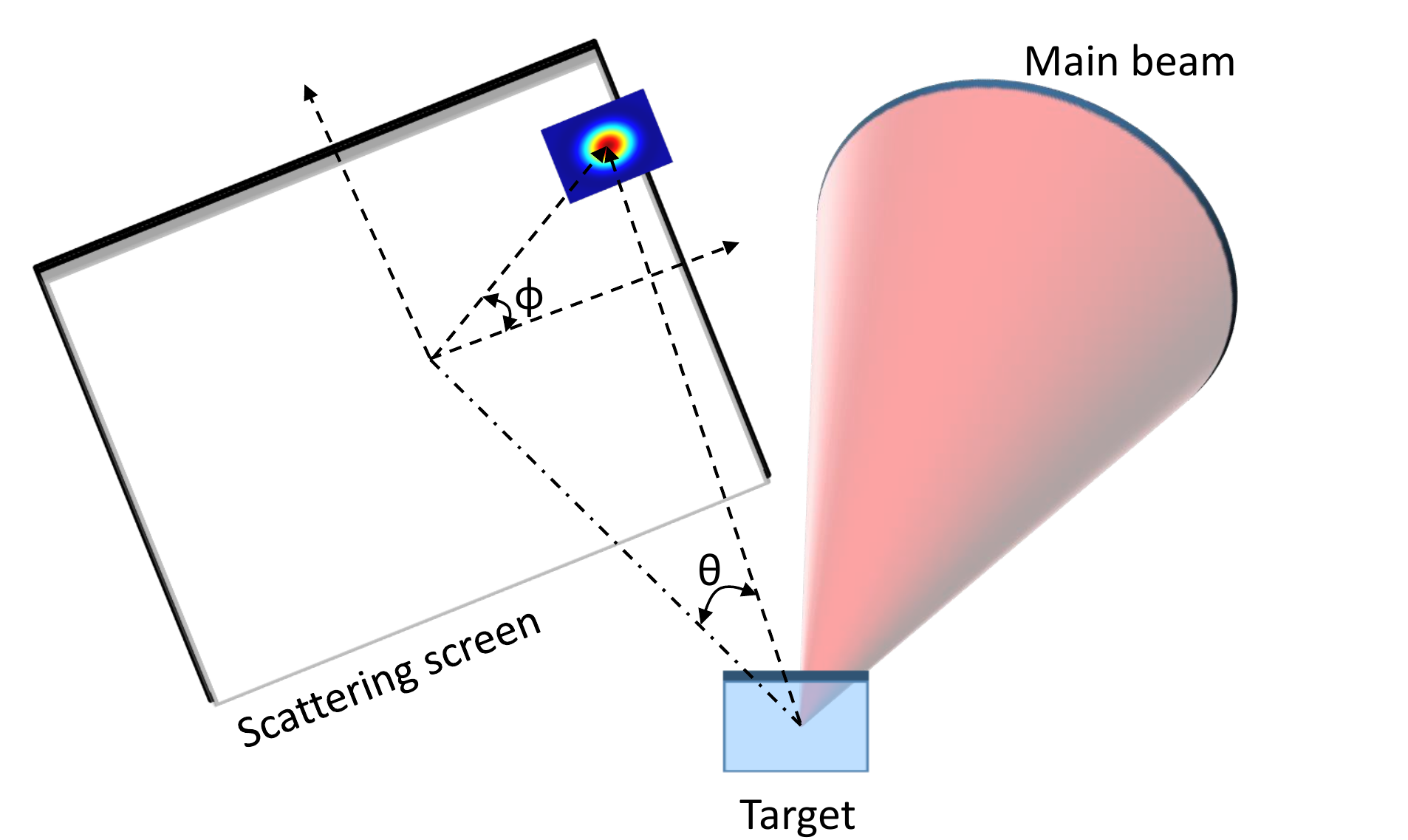}}
\hfill
\subfloat[\label{fig:energy-frac-1w-2w}]{\includegraphics[width=0.5\textwidth, trim=1cm 2cm 2cm 1cm, clip]{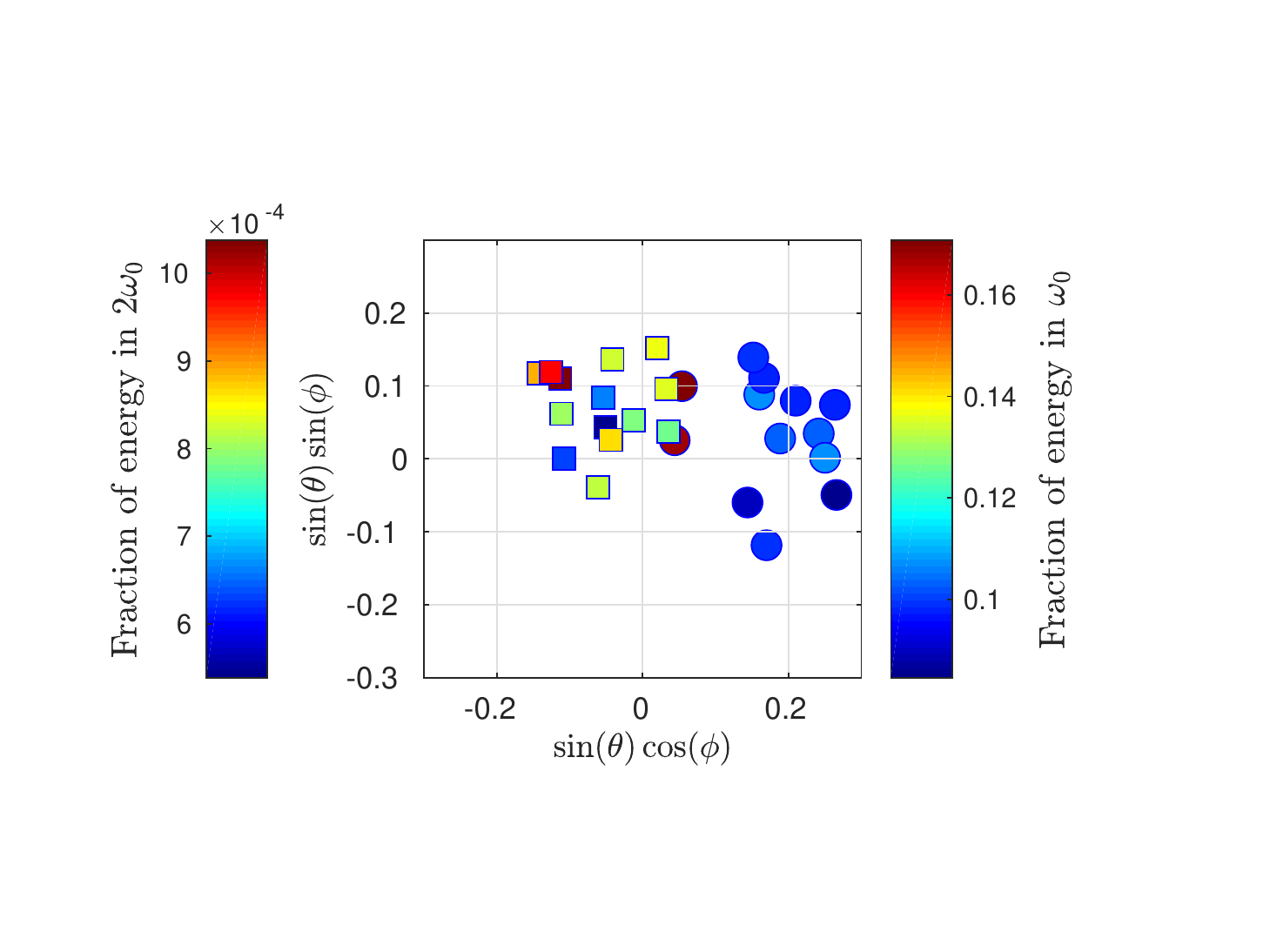}}
\caption{(a) Geometry of the scattering screen monitoring the specular reflection. (b) The fraction of reflected energy on plane of the scattering screen at first (circles) and second harmonic (squares) of the laser frequency. The reflected energy of both frequencies are normalized with laser energy and represented by colorbar in the figure.}
\end{figure}

In order to quantify the deviation of the reflected light at fundamental and second harmonics from the specular direction, we define angles $\theta$ and $\phi$ as shown in figure \ref{fig:scattering-geometry}. The center of the screen is the specular reflection direction and corresponds to $\theta=0$. The angle $\phi$ defines the deviation of the maximum of the scattered radiation with respect to the horizontal plane of laser incidence. Results from the entire experimental campaign are summarized in figure \ref{fig:energy-frac-1w-2w}, which shows fraction of reflected energy on the scattering screen at first (circles) and second harmonic (squares) of the laser frequency. The results indicate that reflected light at first harmonic shifts towards the axis of incoming laser beam while the light generated at second harmonic is centered on the scattering screen. The deviation from the specular direction at fundamental harmonic is $\theta\sim11^\circ$. The circular markers in figure \ref{fig:energy-frac-1w-2w} were derived from the maximum of the part of reflected energy which was incident on the scattering screen. However, as seen in most cases, the maximum was clearly beyond the scattering screen which corresponds to a deviation from specular direction of greater than $17^\circ$. The only two shots for which the reflection of fundamental harmonic was centered on the scattering screen correspond to targets coated with $10\ \mu$m parylene.

\begin{figure}
\centering
\includegraphics[width=0.5\textwidth]{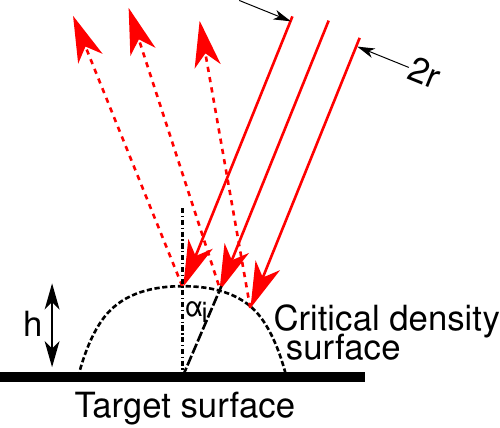}
\caption{\label{fig:schematic-rays-on-critical-surface} Schematic of reflection of incident rays from a curved critical density surface at the focus of a laser beam. The solid red arrows indicate the incident rays of light and the dotted red lines indicate the reflected light from the curved critical density surface. The reflected rays are veered to the right because of the curvature of critical density surface.}
\end{figure}

The fundamental harmonic laser light is reflected from the location where the electron density is equal to $n_c\cos^2\alpha$, where $\alpha$ is the angle the light ray makes with the local density gradient and $n_c=m_e \varepsilon_\circ\left(\frac{2\pi c}{e\lambda}\right)^2$, is the plasma critical density. $m_e$ is the mass of an electron, $\varepsilon_\circ$ is the permittivity of free space, $c$ is the speed of light in vacuum, $e$ is the electronic charge and $\lambda$ is the wavelength of the fundamental harmonic. The consistent bias in the direction of reflection of the fundamental harmonic and not in the second harmonic leads to the following conclusions:
\begin{enumerate}
\item The critical density surface (which is very close to the location where incoming rays are reflected) for the shots with uncoated Ta targets was curved outwards because of plasma expansion initiated by the laser pre-pulse while it was relatively flat for the case of targets coated with plastic. The curved critical density surface for the uncoated target must have expanded at least by a distance $h=r/\tan{\alpha_i}$, where $r$ is radius of the focal spot and $\alpha_i$ is the angle of incidence. This can be explained by a simplified schematic shown in figure \ref{fig:schematic-rays-on-critical-surface}. For the reflected rays (red dotted rays) to be reflected to the right of specular reflection direction, the rays should be incident on critical density surface in the right half of the focal spot. Thus for our experimental parameters of $\alpha_i=22^\circ$, and focal spot radius of $22.5\ \mu$m, the expected height of the critical density surface should be $56\ \mu$m.
\item The relativistically oscillating mirror mechanism is not the likely model applicable for our experiment, because had the second harmonic been generated from relativistic oscillation of the critical density surface, the second harmonic light would have been reflected to the right, similar to the fundamental. Also, as shown below, the density scale lengths expected in the experiment are much larger than the laser wavelength. Thus, neither relativistically oscillating mirror nor coherent wake field emission can be responsible for generating the second harmonic. Instead, for our experiment it is likely that the second harmonic is generated by the mode conversion of the resonant electric field at the critical density surface \cite{Erokhin:PTI:1974}. Such a mode conversion mechanism is applicable to longer plasma density scale lengths. Thus our experiment is different from similar experiments performed with high contrast Ti:Sa lasers where the plasma density profile was very steep and a diffuse emission at the second harmonic was attributed to relativistic oscillations of the critical density surface and originating from the brightest intensities at the center of laser focus \cite{Horlein:ASTRA:2008}.
\end{enumerate}

\begin{figure}
\centering
\includegraphics[width=0.6\textwidth,trim=0cm 2cm 0cm 2cm, clip]{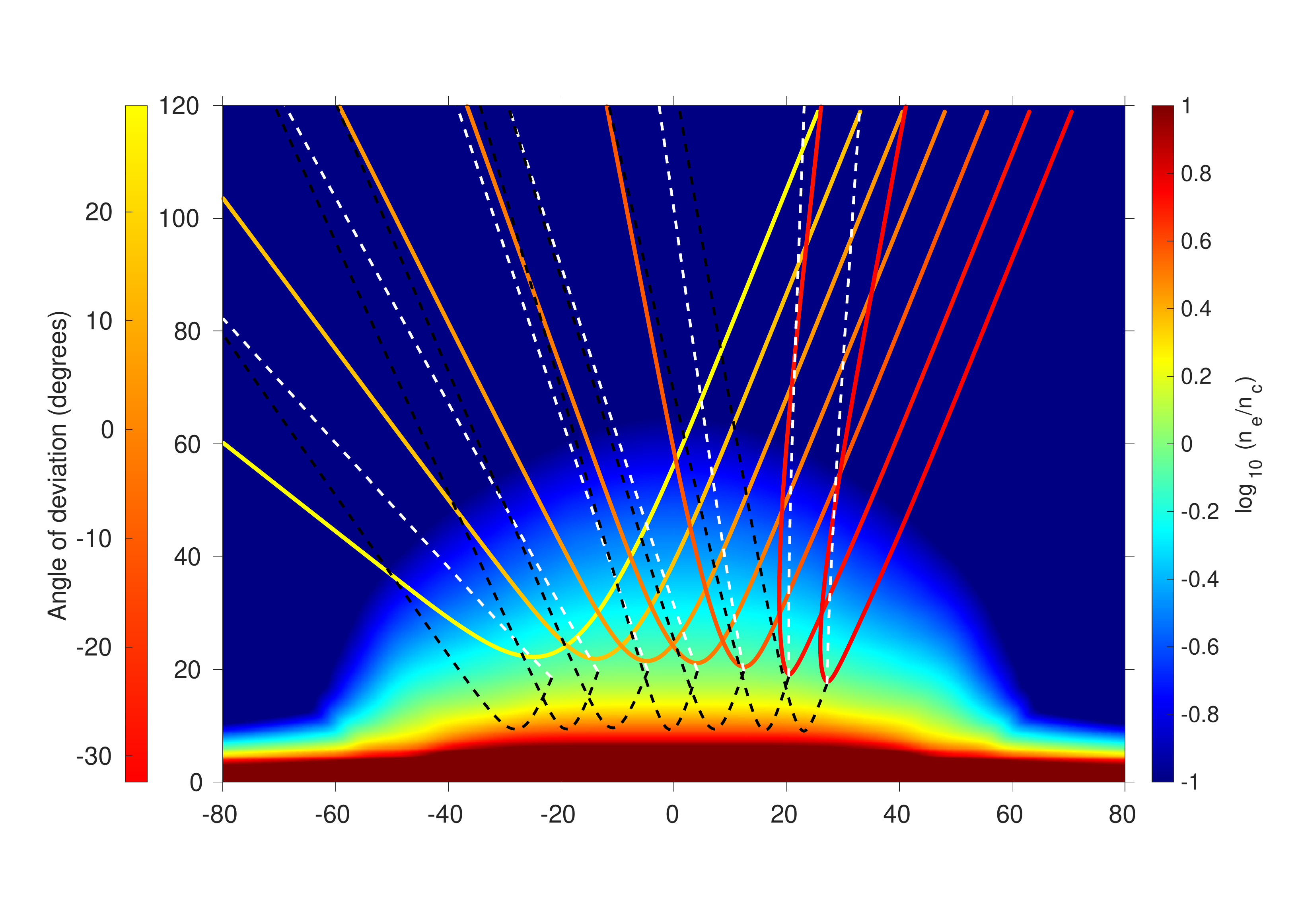}
\caption{\label{fig:hydro-simulations} Simulated electron plasma density for laser pre-pulse incident on Ta targets. The electron density was calculated by assuming that the Ta ions are completely ionized. Colorbar on the right represents the electron density as a fraction of the critical density $n_c$. Solid lines represent the rays at fundamental harmonic launched from the top right corner at an angle of incidence of $22^\circ$. The color of the solid rays (left colorbar) represents that angle of deviation of the reflected outgoing ray, where $0^\circ$ corresponds to direction of specular reflection from a mirror. The dashed lines represent the propagation of the rays at the second harmonic which are launched from the critical density surface. The white lines propagate outwards, while the black rays propagate towards the target before being reflected near the plasma with electron density $\sim 4n_c$. Distance units in $\mu$m.}
\end{figure}

Two dimensional hydrodynamic simulations with cylindrical symmetry were performed using the FLASH code \cite{Fryxell:FLASH:2000} to estimate the effect of pre-pulse on the expansion of target material and formation of the critical density surface. The code uses arbitrary mesh refinement of a finite volume Eulerian grid and includes ray tracing model of laser energy deposition. The code also includes electron and radiation energy transport and uses separate equations for electron, ion and radiation temperatures. The equations of state and opacities of tantalum, carbon and hydrogen were calculated from the QEOS model \cite{More:QEOS:1988}. Simulations were performed in the two-dimensional axi-symmetric geometry. The laser power corresponding to the ASE and the pedestal shown in figure \ref{fig:laser-contrast} was focused on a $50\ \mu$m diameter spot. For the simulations, the incident ASE power level was extended to $850$ ps before the main pulse. Laser energy was absorbed via inverse bremsstrahlung, and so the code was only able to simulate ablation and plasma expansion up to the time $t_2$ at $10$ ps before the main pulse (see figure \ref{fig:laser-contrast}), where the focused intensity was less than $10^{15}$ W/cm$^2$. The last $10$ ps before the main pulse cannot be simulated with a hydrodynamic code, but this has negligible effect on the plasma density profile as the velocity of expanding plasma is of the order of $0.1\ \mu$m/ps.

Simulations were performed for tantalum (plastic coated tantalum) targets, and predicted a corona electron temperature of $\sim85$ eV ($\sim 100$ eV) during the ASE, consistent with the laser ablation model \cite{Atzeni:INFM:2004}. During the ASE (before time $t_1$ in figure \ref{fig:laser-contrast}), the critical density surface expands to a distance of $\sim7\ \mu$m ($\sim3\ \mu$m) from the target surface. Thus, the critical density surface is planar and agrees with the observed specular reflection from the plastic coated targets. However, it cannot explain the observed shift in the reflection of fundamental harmonic from uncoated targets. Even the two orders of magnitude increase in intensity during the pedestal (between the times $t_1$ and $t_2$) was insufficient to significantly alter the critical density surface as the duration of pedestal is only $\sim 50$ ps.

The possible explanation of the experimentally observed shift in fundamental harmonic invokes photo-ionization of the Ta plasma by the main laser pulse. While carbon ions (Z$=6$) are fully stripped in the pre-plasma, charge of tantalum ions (Z=73) is $\sim 14$. Therefore, the main pulse of relativistic intensity may increase the electron density in the corona almost instantaneously by a factor of $5$. To estimate the effect photo-ionization might have on the critical density surface, we use the ion density profile from the hydrodynamic simulation for Ta target and multiply it with an expected maximum ion charge of $73$ to get the electron density profile. This can be expected as the ionization energy of Ta$^{72+}$ is $\approx 70$ keV. The resulting increase in electron density moves the surface of reflection to $22\ \mu$m, as can be seen in figure \ref{fig:hydro-simulations}. The effect of such a density profile on the reflection of incoming rays and the direction of propagation of second harmonic is also shown in figure \ref{fig:hydro-simulations}. For imitating the propagation of light within the Rayleigh length near the focus, individual rays at fundamental harmonic were launched from the top right corner at an angle of $22^\circ$ within a diameter of $45\ \mu$m, corresponding to the focal spot in the experiment. As seen in figure \ref{fig:hydro-simulations}, most of the rays are reflected away from the specular direction towards the incoming laser axis (i.e., corresponding to a negative angle of deviation) as observed in the experiment. Generation of the second harmonic is described according to the theoretical model by Erokhin \etal \cite{Erokhin:PTI:1974}. As shown in figure \ref{fig:hydro-simulations} with dashed lines, these rays propagate such that they preserve the transverse component of the momentum of the incoming rays at fundamental frequency. The second harmonic rays are less refracted compared to the fundamental frequency and show a more diffuse pattern similar to the experimental observation. The plasma density profile and the ray propagation as shown in figure \ref{fig:hydro-simulations} provide only a qualitative illustration that photo ionization of tantalum can provide a significantly curved critical density surface which can explain the measured deviation of the reflected light in the experiment. A consistent calculation of the density profile by including photo-ionization of tantalum and the corresponding light propagation and generation of second harmonic is beyond the scope of the paper.

The energy measured at the second harmonic depends on the target material. The reflected energy in the range of $0.2 - 0.3$ J was measured at the second harmonic for uncoated Ta targets, while the emission from plastic coated targets was about $30\%$ lower. A higher emission at the second harmonic for targets having high atomic number was also reported by Raffestin \etal \cite{Raffestin:CELIA:2019}. The measured conversion efficiency of $\lesssim0.1\%$ to the second harmonic was significantly less than the conversion efficiency of $\sim 10\%$ reported for Ti:Sa lasers with similar focused intensity \cite{Streeter:NJP:2011}. The conversion efficiency of laser energy reported here into second harmonic is a lower bound because some radiation falls beyond the scattering screen. The measured conversion efficiency provides an estimation for the electron density profile near the critical density. According to the theoretical model \cite{Erokhin:PTI:1974}, the conversion efficiency $Q_{2\omega}$ depends on three parameters: the angle of incidence $\alpha_i$, the ratio of the density scale length to the laser wavelength $\rho = 2\pi L_n/\lambda$, and on the dimensionless laser amplitude $a_0= eE_0/m_e \omega c$, as follows
\begin{equation}
Q_{2\omega} \sim a_0^2 \rho^2 \sin^2\alpha_i Q_{res}^2,
\end{equation}
where $Q_{res}\sim\rho^{2/3} \sin^2\alpha_i e^{-4\rho \sin^3\alpha_i/3}$ is the efficiency of resonance absorption of the laser near the critical density. The function $Q_{res}$ for our experimental parameters strongly depends on plasma density scale length, it decreases from $\sim10^{-2}$ to $\sim10^{-5}$ for $\rho$ increasing from $100$ to $200$. Similarly, assuming $a_0\sim2-3$ for the main pulse, the efficiency of second harmonic emission decreases from $0.1$ to $10^{-5}$. The measured efficiency of $10^{-3}$ is within this range. We thus conclude that the plasma density scale length in the experiment is in the range of $20-30\ \mu$m.

\subsection{\label{sec:characterization of energetic electrons} Characterization of energetic electrons}
\begin{figure}
\centering
\includegraphics[width=0.5\textwidth]{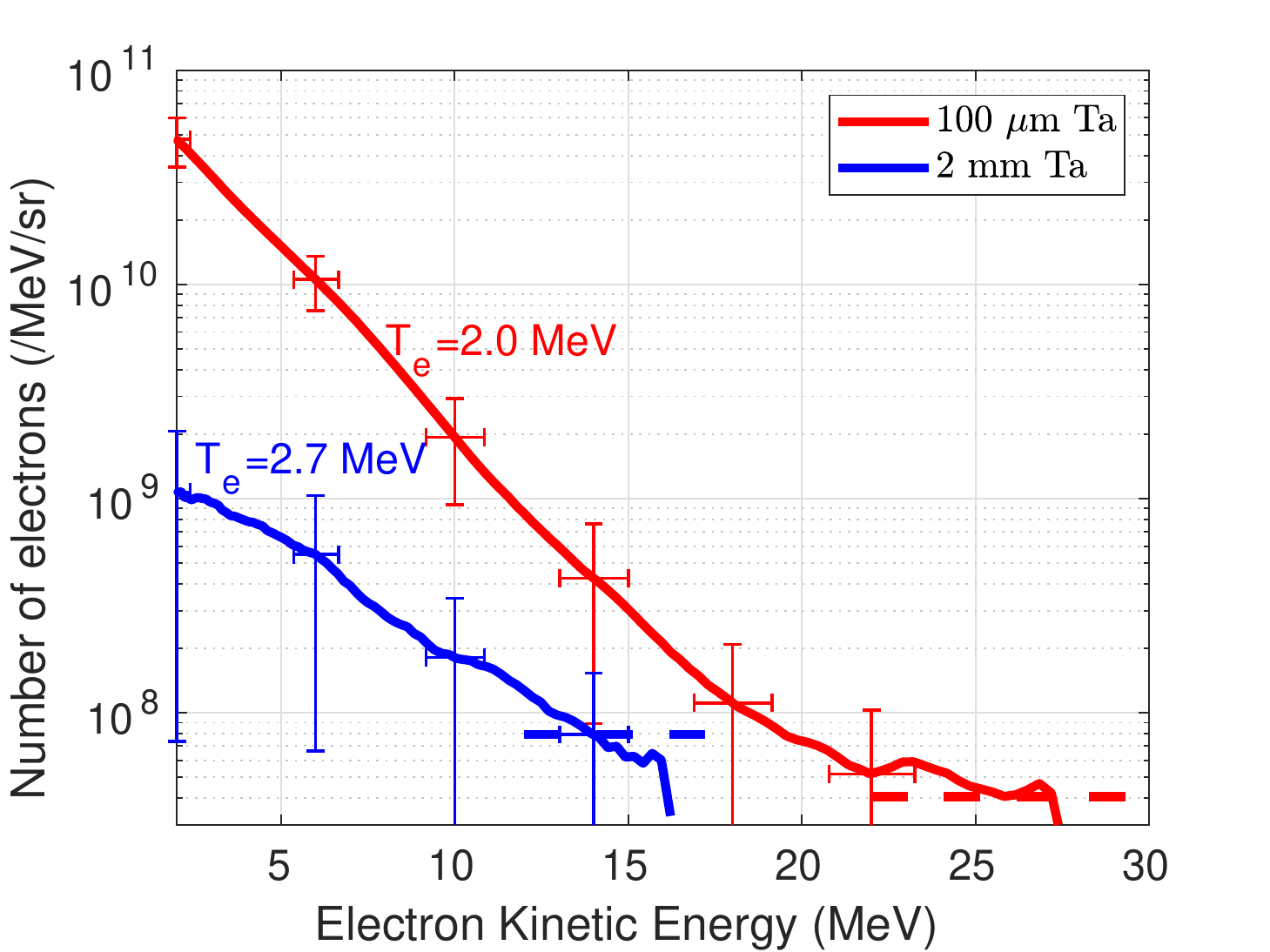}
\caption{\label{fig:electron-spectrum} Experimental measurements of escaped electrons energy distribution from a single shot for Ta target of thickness of $0.1$ mm (red line) and $2$ mm (blue line). The dotted lines represent the corresponding noise level in the spectra.}
\end{figure}

The energy spectrum of escaped electrons was measured directly by a magnetic spectrometer installed at a distance of $20$ cm from the Ta target (see figure \ref{fig:schematic-layout-sgii}). Figure \ref{fig:electron-spectrum} shows the measured electron spectrum for $100\ \mu$m and $2$ mm thick targets. The electron temperature estimated from the slope of the distribution is $(2\pm0.2)$ and $(2.7\pm0.3)$ MeV respectively. The total energy of escaping electrons can be estimated assuming the solid angle of emission of $0.4$ sr, as it was measured in other similar experiments \cite{Rusby:GSI:2018}. Then, for the 100 $\mu$m thick target, a total number of electrons with energy greater than $1$ MeV escaping the target is of the order of $10^{11}$, which corresponds to a charge of about $20$ nC. The total energy carried by these electrons is $\sim 0.05$ J, indicating a conversion efficiency from laser energy to escaped electrons of the order of $10^{-4}$.

The electron temperature measured from laser interaction with a $100\ \mu$m thick target is a factor of two higher than the ponderomotive scaling \cite{Wilks:LLNL:2001}, assuming the dimensionless laser field amplitude $a_0=2-3$, but similar to measurements on other high energy Nd:glass laser systems \cite{Tanimoto:Osaka:2009,Yogo:LFEX:2017,Williams:NIF:2020} and also predicted by simulations \cite{Cui:BNL:2013}. This is explained by contribution of the direct laser acceleration and stochastic heating in a plasma extended to $10$ laser wavelengths or more \cite{Williams:NIF:2020}.

The number of electrons detected from $2$ mm thick targets is reduced by a factor of about $50$ and the maximum cutoff energy of electrons is decreased by about $6$ MeV. This is consistent with the expected energy lost by fast electrons while traversing $2$ mm thick tantalum targets \cite{STAR:NIST}. The lower energy electrons are scattered more within the target compared to the high energy electrons. Thus, a higher fraction of high energy electrons are able to escape the thick target and reach the electron spectrometer compared to the low energy ones. This results in an apparent increase in hot electron temperature measured by the spectrometer for thick targets, but it is only an artifact of electron scattering in thicker targets and not a result of higher electron temperatures at the laser focus. The electron spectrometer data thus shows that electron temperature about $2$ MeV and energies up to $20$ MeV were generated in the laser plasma interaction.

\subsection{\label{sec:Hard X-ray characterization} Hard X-ray characterization}
\begin{figure}
\subfloat[\label{fig:scintillator-lineouts}]{\includegraphics[width=0.33\textwidth,trim=0.5cm 0cm 0cm 0cm, clip]{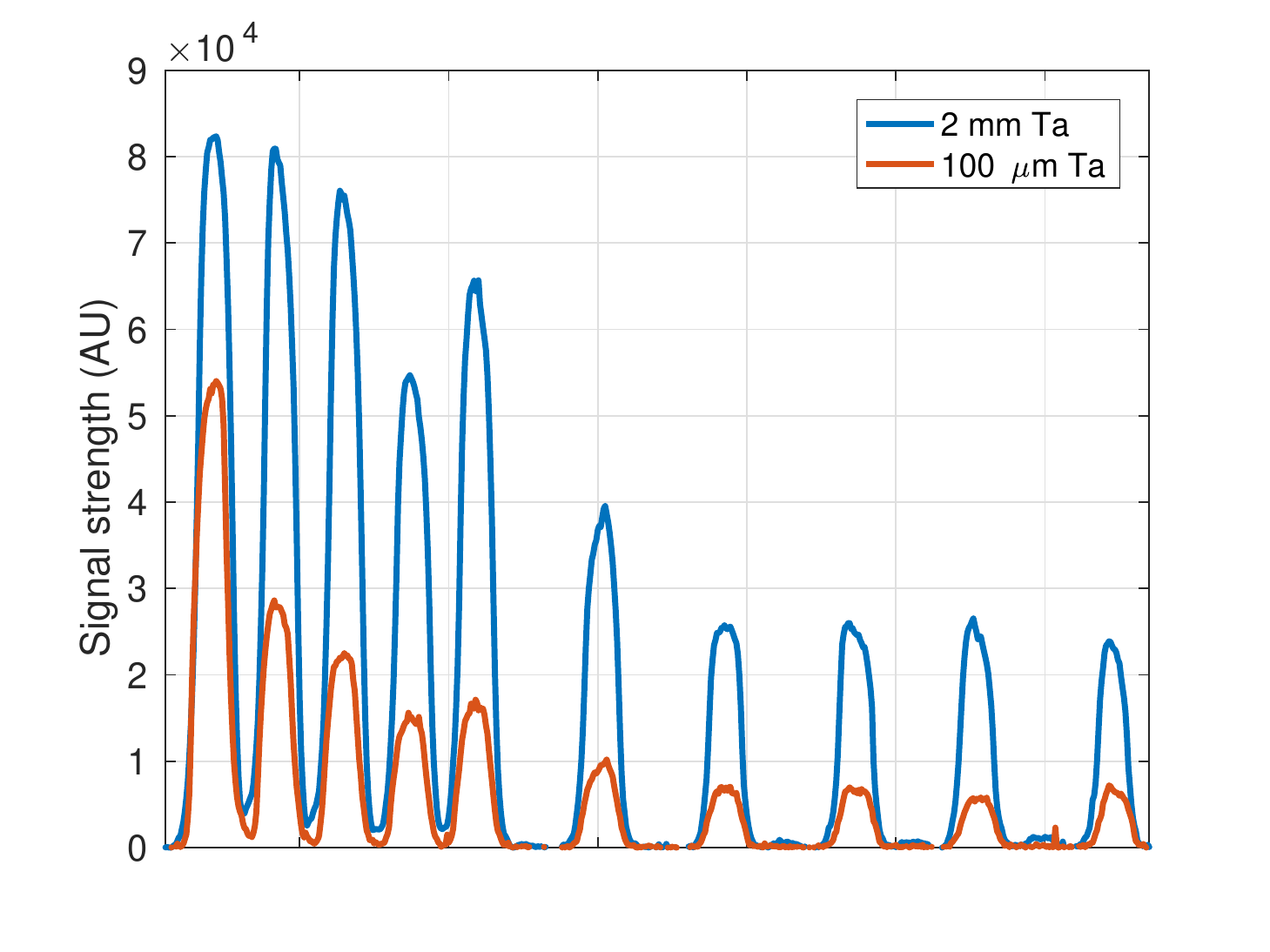}}
\hfill
\subfloat[\label{fig:MC-spectra}]{\includegraphics[width=0.33\textwidth,trim=0cm 0cm 0cm 0cm, clip]{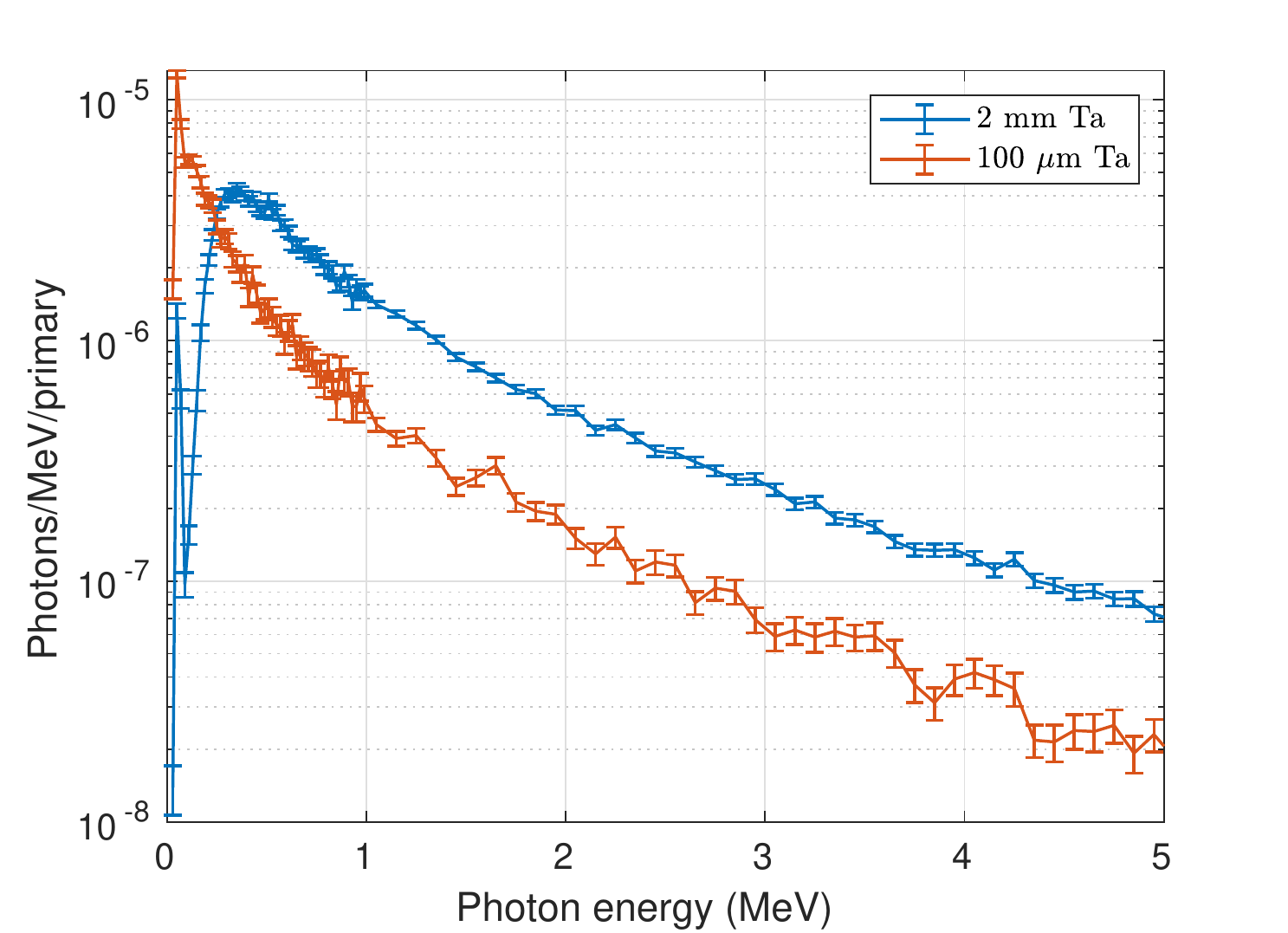}}
\hfill
\subfloat[\label{fig:MC-comparison}]{\includegraphics[width=0.33\textwidth,trim=0.5cm 0cm 0cm 0cm, clip]{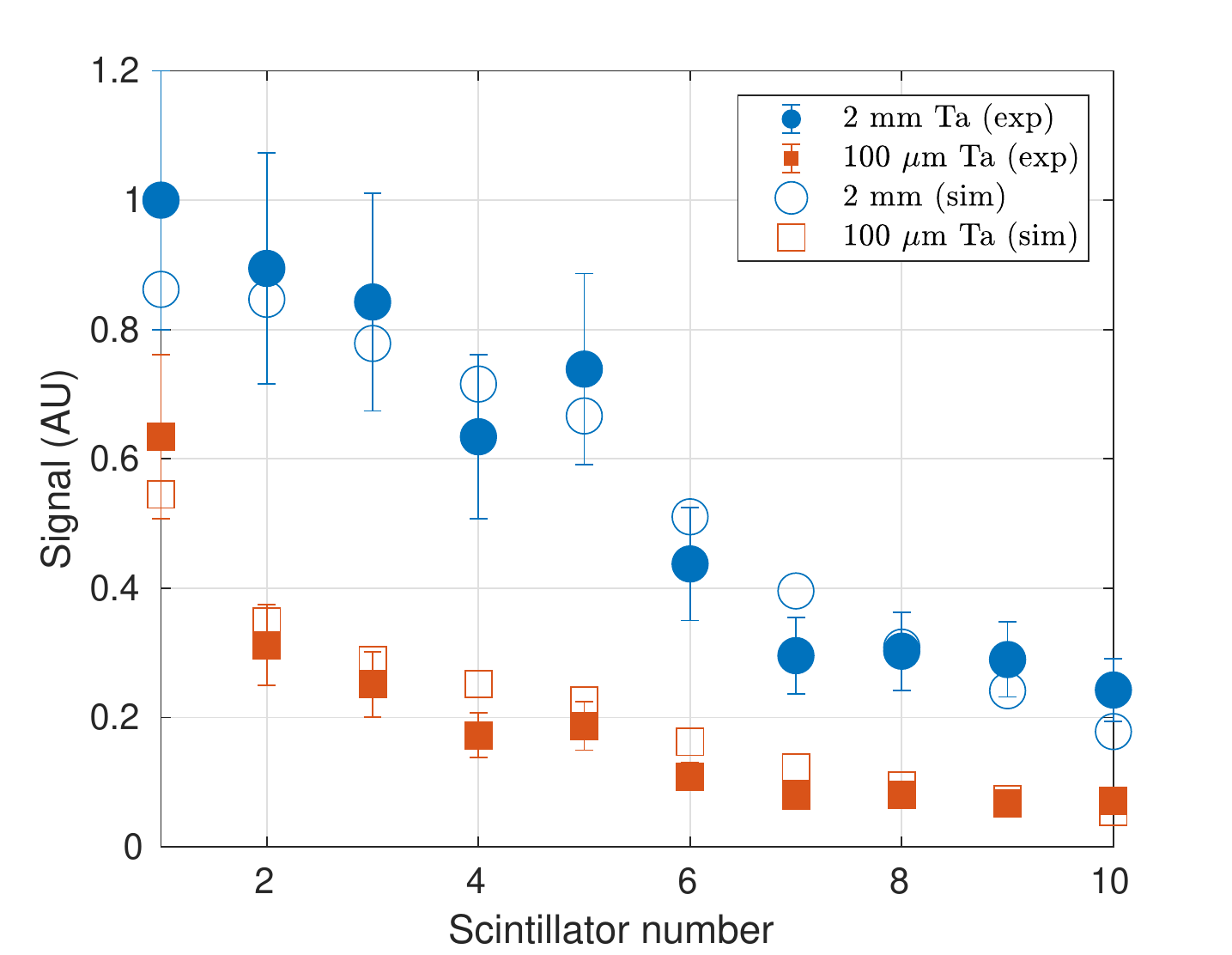}}
\caption{(a) Average brightness across the data from scintillator stack. The ten peaks correspond to the ten scintillators, with the peak on the left corresponding to low energy photons. (b) Expected photon spectra from the experiment generated by Monte Carlo simulations. (c) Comparison of the predicted response of the scintillator stack to the measurement.}
\label{fig:xray-data}
\end{figure}
X-ray measurements from targets of two different thicknesses are shown in figure \ref{fig:scintillator-lineouts}. The spectrum  from the $100\ \mu$m thick Ta target is dominated by low energy photons of energy $100-200$ keV, as evident from the significantly lower signal in the second scintillator compared to the first. These photons are generated by a low energy, non-relativistic component of electrons in plasma. Their bremsstrahlung emissivity depends inversely on the electron energy \cite{Lamoureux::2006} and the target thickness is comparable to their stopping range \cite{STAR:NIST}. The emissivity of relativistic electrons is much smaller and the corresponding high energy photons deposit their energy in all the scintillators in the stack (see also figure \ref{fig:scintillator-stack-energy-deposited}), but the energy resolution as defined by the first few layers is insufficient to deduce their energy distribution. The higher intensity of the data collected from $2$ mm thick targets and the relatively gradual decay along the scintillators (see figure \ref{fig:scintillator-lineouts}) implies that the photon spectrum from the $2$ mm thick Ta target is dominated by harder photons and also has a higher flux.

These observations are supported with dedicated Monte Carlo simulations for expected photon spectra shown in figure \ref{fig:MC-spectra}. The simulations were performed with the FLUKA code \cite{FLUKA:website,Battistoni:FLUKA:2015,Bohlen:FLUKA:2014} and used the electron spectra shown in figure \ref{fig:electron-spectrum} as an input. The electrons were injected along the direction of laser propagation and had an opening cone angle of $\pm15^\circ$ \cite{Daykin:POP:2018,Sawada:Leopard:2019}. The temperature of the photon distribution for energies greater than $1$ MeV was $T_\gamma\sim0.7$ MeV, i.e., of the same order as the electron temperature for both the targets \cite{Palaniyappan:LANL:2018}. However, the low energy photons with energy less than $\sim150$ keV are significantly attenuated within the target \cite{XCOM:NIST}.

The difference in the characteristic bremsstrahlung spectra from the different kinds of targets make them useful for contrasting applications. Thin targets are used for generating X-rays to probe high energy density physics experiments where X-rays with energies of few tens of keV are required \cite{Sawada:Leopard:2019}. Thick targets providing photons with energy of several MeV are used for applications related to photo-nuclear activation and for measuring cross section for transmutation of waste products \cite{Ledingham:Science:2003}. From the measured brightness of the scintillators, we expect a target thickness of $2-3$ mm to be optimal for bremsstrahlung production for photons with energy of few MeV \cite{Galy_2007}.

The spectra from the Monte Carlo simulations shown in figure \ref{fig:MC-spectra} were convolved with the response matrix of the scintillator stack shown in figure \ref{fig:scintillator-stack-energy-deposited} to predict the response of the diagnostic. Only one normalization constant was adjusted to minimize the Chi-squared residue for fitting the spectra from both targets. A good agreement with the experimental measurement can be seen in figure \ref{fig:MC-comparison}. This further confirms that the hot electrons with a mean energy of $2$ MeV is the dominant component of laser accelerated electrons.

\section{\label{sec:conclusion}Conclusion}
In summary, this paper presents results from an experiment in which a short pulse from a Nd:glass laser was focused on Ta target of thickness ranging from $100\ \mu$m to $4$ mm at an intensity of $10^{19}$ W/cm$^2$. Measurements of the optical emission in the specular reflection direction provide information about a pre-plasma density profile which extends to several tens of $\mu$m because of the photo-ionization of expanding tantalum plasma. The amount of energy specularly reflected in the fundamental harmonic is as high as $25\%$. Efficiency of the laser energy conversion into second harmonic confirms the estimated pre-plasma scale length of $20-30\ \mu$m.

The hot electrons are characterized by measuring the energy spectrum for $100\ \mu$m and $2$ mm thick Ta targets. The measured electron temperature of $2$ MeV and dependence of the bremsstrahlung photon yield on the target thickness are in close agreement with results from Monte Carlo simulations. These results are important for development of new efficient photon sources and for designing hot electron diagnostic methods in relativistic laser plasma interactions. Previous experiments reported up to a two-fold enhancement in X-rays from targets coated with plastic \cite{Courtois:POP:2009}. However, we only measured about a $25\%$ increase, which is within the uncertainty limit due to shot-to-shot fluctuation.

\section*{Acknowledgments}
This research was sponsored by the Czech Science Foundation (project No. 18-09560S) and by the project High Field Initiative (CZ$.02.1.01\/0.0\/0.0\/15\_003\/0000449$) from the European Regional Development Fund (HIFI). The results of the project LQ1606 were also obtained with the financial support of the Ministry of Education, Youth and Sports as part of targeted support from the National Programme of Sustainability II. The work was also supported by the project Advanced research using high intensity laser produced photons and particles (CZ$.02.1.01\/0.0\/0.0\/16\_019/0000789$) from European Regional Development Fund (ADONIS).

\bibliographystyle{iopart-num}
\bibliography{sgii}

\end{document}